\documentclass[journal=nalefd,manuscript=letter,layout=preprint]{achemso}

\usepackage{amsmath,graphicx,latexsym,times,color}
\usepackage{setspace}
\usepackage{hyperref}
\usepackage{array}
\usepackage{titlesec}
\usepackage{physics}
\usepackage{color}
\usepackage[T1]{fontenc}
\usepackage[version=3]{mhchem} 

\usepackage{empheq} 

\author{Dongzhe Li}
\email{dongzhe.li@cemes.fr}
\affiliation{Department of Physics, Technical University of Denmark, DK-2800 Kongens Lyngby, Denmark}
\alsoaffiliation{CEMES, Universit\'e de Toulouse, CNRS, 29 rue Jeanne Marvig, F-31055 Toulouse, France}

\author{Yongfeng Tong}
\affiliation{Universit\'e Paris Cit\'e, CNRS, Laboratoire Matériaux et Phénomènes Quantiques UMR7162, 75013 Paris, France}

\author{Kaushik Bairagi}
\affiliation{Universit\'e Paris Cit\'e, CNRS, Laboratoire Matériaux et Phénomènes Quantiques UMR7162, 75013 Paris, France}

\author{Massine Kelai}
\affiliation{Universit\'e Paris Cit\'e, CNRS, Laboratoire Matériaux et Phénomènes Quantiques UMR7162, 75013 Paris, France}

\author{Yannick J. Dappe}
\affiliation{Universit\'e Paris-Saclay, CEA, CNRS, SPEC, 91191 Gif-sur-Yvette, France}

\author{Jérôme Lagoute}
\affiliation{Universit\'e Paris Cit\'e, CNRS, Laboratoire Matériaux et Phénomènes Quantiques UMR7162, 75013 Paris, France}

\author{Yann Girard}
\affiliation{Universit\'e Paris Cit\'e, CNRS, Laboratoire Matériaux et Phénomènes Quantiques UMR7162, 75013 Paris, France}

\author{Sylvie Rousset}
\affiliation{Universit\'e Paris Cit\'e, CNRS, Laboratoire Matériaux et Phénomènes Quantiques UMR7162, 75013 Paris, France}

\author{Vincent Repain}
\affiliation{Universit\'e Paris Cit\'e, CNRS, Laboratoire Matériaux et Phénomènes Quantiques UMR7162, 75013 Paris, France}

\author{Cyrille Barreteau}
\affiliation{Universit\'e Paris-Saclay, CEA, CNRS, SPEC, 91191 Gif-sur-Yvette, France}

\author{Mads Brandbyge}
\affiliation{Department of Physics, Technical University of Denmark, DK-2800 Kongens Lyngby, Denmark}
\alsoaffiliation{Center for Nanostructured Graphene, Technical University of Denmark, DK-2800 Kongens Lyngby, Denmark}

\author{Alexander Smogunov}
\affiliation{Universit\'e Paris-Saclay, CEA, CNRS, SPEC, 91191 Gif-sur-Yvette, France}

\author{Amandine Bellec}
\email{amandine.bellec@u-paris.fr}
\affiliation{Universit\'e Paris Cit\'e, CNRS, Laboratoire Matériaux et Phénomènes Quantiques UMR7162, 75013 Paris, France}

\title[\texttt{achemso} demonstration]
{Negative Differential Resistance in Spin-Crossover Molecular Devices}

\let\oldtimes\times  
\renewcommand\times{{\oldtimes}}

\begin{document}

\begin{tocentry}
	\begin{center}
		\includegraphics[width=1.0\linewidth]{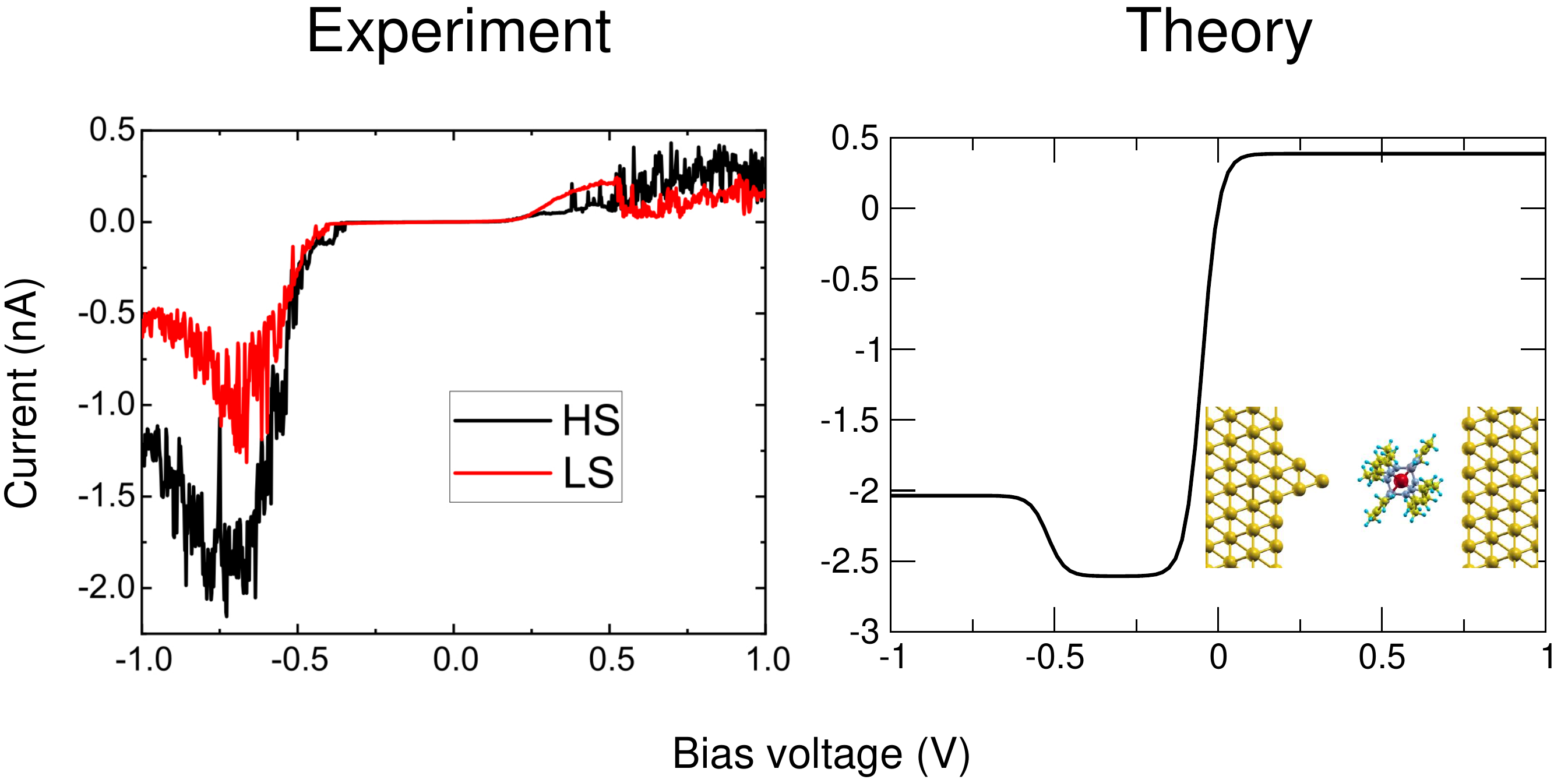}
	\end{center}
\end{tocentry}
	
	\begin{abstract}
		We demonstrate, based on low-temperature scanning tunneling microscopy (STM) and spectroscopy, a pronounced negative differential resistance (NDR) in spin-crossover (SCO) molecular devices, where a Fe$^{\text{II}}$ SCO molecule is deposited on surfaces. The STM measurements reveal that the NDR is robust with respect to substrate materials, temperature, and the number of SCO layers. This indicates that the NDR is intrinsically related to the electronic structure of the SCO molecule. Experimental results are supported by density functional theory (DFT) with non-equilibrium Green's functions (NEGF) calculations and a generic theoretical model. While the DFT+NEGF calculations reproduce NDR for a special atomically-sharp STM tip, the effect is attributed to the energy-dependent tip density of states rather than the molecule itself. We, therefore, propose a Coulomb blockade model involving three molecular orbitals with very different spatial localization as suggested by the molecular electronic structure. \\
		Keywords: Negative differential resistance, Coulomb blockade, scanning tunneling microscopy, spin crossover molecule, density functional theory, Landauer-B\"uttiker scattering theory
	\end{abstract}
	
	Spin-crossover (SCO) molecules, a family of molecular bistable systems with easy spin manipulation via external stimuli, have attracted tremendous attention due to their potential application in molecule-based spintronics devices \cite{ali2012electronic,bousseksou2011molecular, Lefter2016, Molnar2018, Zhang2020_device}. Utilizing single SCO molecules as quantum units opens a new pathway to reach the physical limits of miniaturization and achieve energy-efficient control and detection of magnetism \cite{gruber2020spin,montenegro2021deposition,kipgen2021spin,konstantinov2021electrical,torres2021spin,boix2022strain,Gavara-Edo2022}. In this context, a fundamental understanding of the current-voltage ($I-V$) behavior is mandatory for the smart design of future organic devices. In particular, negative differential resistance (NDR), characterized by a decreasing current with increasing bias in a certain bias voltage window, is one of the most attractive quantum tunneling phenomena. Stimulated by the promising electronic applications that the NDR effect can offer, the functional $I-V$ behavior entailed in the NDR of single-molecule junctions has been extensively studied during the last two decades \cite{Xue1999,mujica2001current,Galperin2005,Muralidharan2007,Chen2007,kang2010observation,Heinrich2011,perrin2014large,xu2015negative,warner2015tunable,tawfik2017near,Kuang2018,Fung2019,Jiang2019,liu2020sharp,Mengting2021}.
	
	Several NDR mechanisms have been proposed by quantum transport theory and model analysis, namely resonances in the projected density of states (PDOS) of the scanning tunneling microscopy (STM) tip  \cite{Xue1999}, voltage drop along the molecule \cite{mujica2001current}, electron-phonon interaction \cite{Galperin2005,liu2020sharp}, Coulomb blocking \cite{Muralidharan2007}, surface states \cite{Heinrich2011}, and orbital symmetry matching \cite{Chen2007}. However, most reported NDR effects are restricted to nonmagnetic molecular systems and have not yet been demonstrated by experiments in SCO systems. The NDR associated with magnetic orderings may give rise to all-electrical spin manipulation due to versatile control of the SCO complex via external stimuli. Moreover, while direct comparison between theory and experiments has become standard for conductance, thermopower, and Fano-factor studies in molecular junctions, NDR studies in corresponding molecular junctions are still missing.
	
	In the present Letter, we report the direct experimental observation of pronounced NDR in the current-voltage characteristics of SCO molecule, Fe$^{\mathrm{II}}$((3,5-(CH$_{3}$)$_{2}$Pz)$_{3}$BH)$_{2}$ (named FeMPz in the following, Pz refers to pyrazolyl), deposited on various surfaces. Interestingly, the pronounced NDR effect appearing for negative bias is robust with respect to substrate materials (ranging from metal to semi-metal), temperature, and the number of SCO layers. These results are analyzed and explained by density functional theory (DFT) with non-equilibrium Green's functions (NEGF) calculations and a Coulomb blockade model. Our calculations conclude on two main points: (i) The mean-field DFT+NEGF approach fails to reproduce molecular NDR, presumably due to the lack of strong correlation effects; (ii) The simple model, including Coulomb interactions on the molecule, reproduces experimental observations, tracing the origin to a specific strongly localized molecular orbital blocking electron transport through the other (more delocalized) orbitals.
	
\begin{figure*}[!h]
		\includegraphics[width=1.0\linewidth]{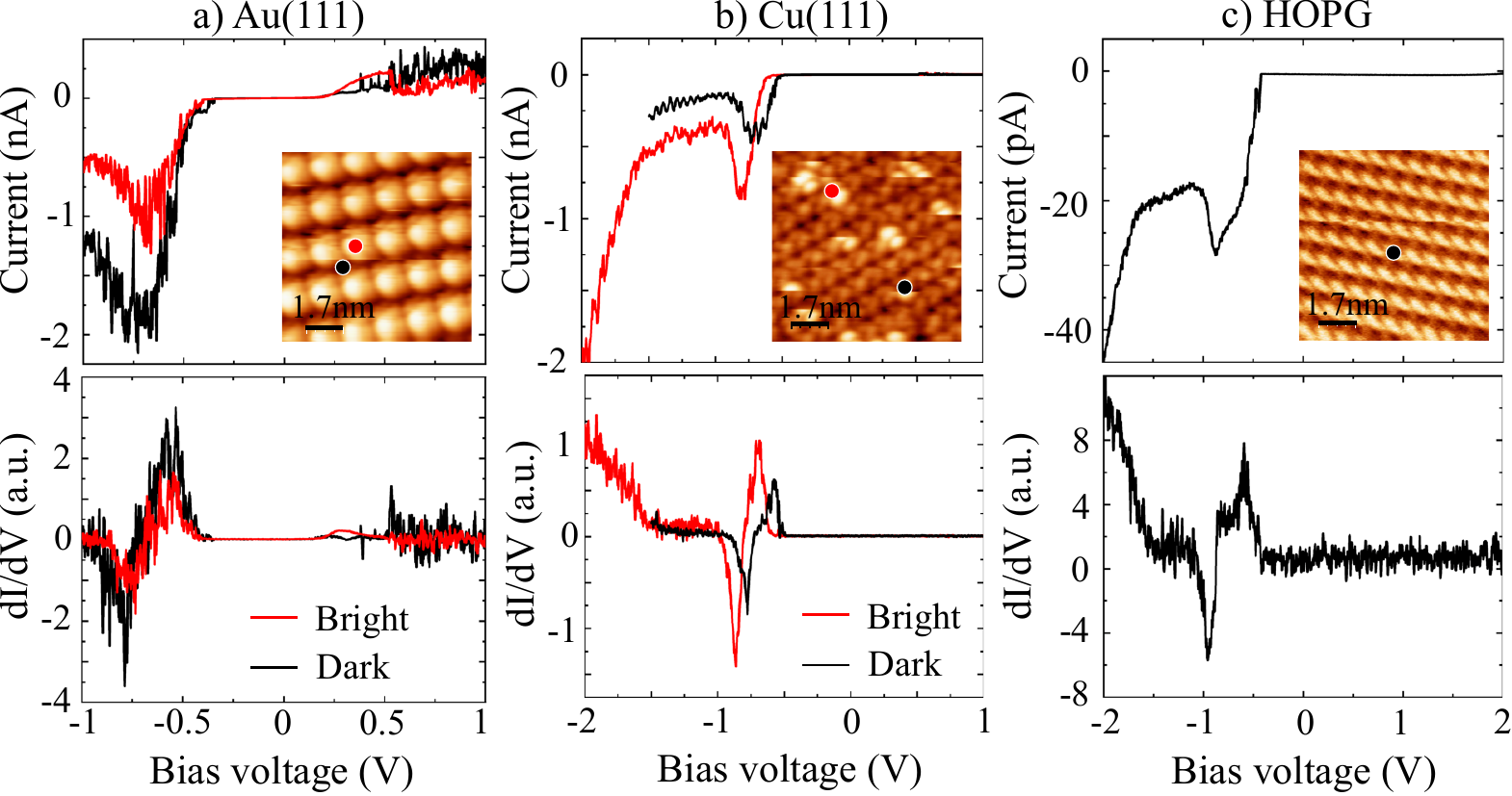}
		\caption{\label{STS_mono}
		STS measurements on molecules in monolayer islands grown on a) Au(111), b) Cu(111), and c) HOPG substrates. For a) and b), the $I-V$ and $dI/dV$ curves have been acquired on molecules appearing either dark or bright at 0.3~V. The 8.5$\times$8.5~nm$^2$ STM topographic images with the positions at which the spectra have been acquired are shown in the inset. a)   0.3~V, 50~pA, b) 0.3~V, 5~pA and c) $-1.5$~V, 20~pA.		
		}
	\end{figure*}

	As reported in the literature for sub-monolayer coverage, the FeMPz molecules self-assemble on various substrates forming one-monolayer-high islands. Interestingly, using a bias voltage of 0.3~V to acquire STM images (inset of Fig. \ref{STS_mono}a-b), two types of molecules (bright and dark) are identified, which can be correlated to the spin state of the molecules \cite{Bairagi2016, Tong2021}. Indeed, the thermal transition is partial, and at low temperature both spin states co-exist in the molecular network \cite{Kelai2021a}. To investigate the electronic properties of the FeMPz molecules, we performed scanning tunneling spectroscopy (STS) measurements.  Fig.~\ref{STS_mono} presents typical $I-V$ and $dI/dV$ curves recorded over FeMPz molecules on Au(111), Cu(111) and highly ordered pyrolytic graphite (HOPG). At positive voltages, on bright molecules, a molecular state can be measured at low bias voltage (Fig.~\ref{STS_mono}a), thus explaining the contrast observed at 0.3~V. The exact energy position of the molecular orbital cannot be determined as the molecules are switching between both spin states for bias voltage above 0.5~V as evidenced by the sharp transition in the $I-V$ curves \cite{Tong2021}. No other molecular state is detected for positive bias. At negative bias, and for all substrates investigated, the STS curves exhibit a clear NDR whose position can vary between 0.6 and 1.1~V. The NDR energy position can not directly be related to the spin state. Indeed, the difference observed in Fig.~\ref{STS_mono}b on Cu(111) could originate from the tip distance being more important over the bright molecule compared to the dark one during the spectra acquisition. Moreover, we find that the NDR signature is robust with temperature. As visible in Fig. S1 in Supporting Information (SI), at liquid N$_2$ temperature, the NDR is observed both over the bright and dark molecules when adsorbed on Au(111).
	
\begin{figure}[!h]
		\centering
		\includegraphics[width=0.9\linewidth]{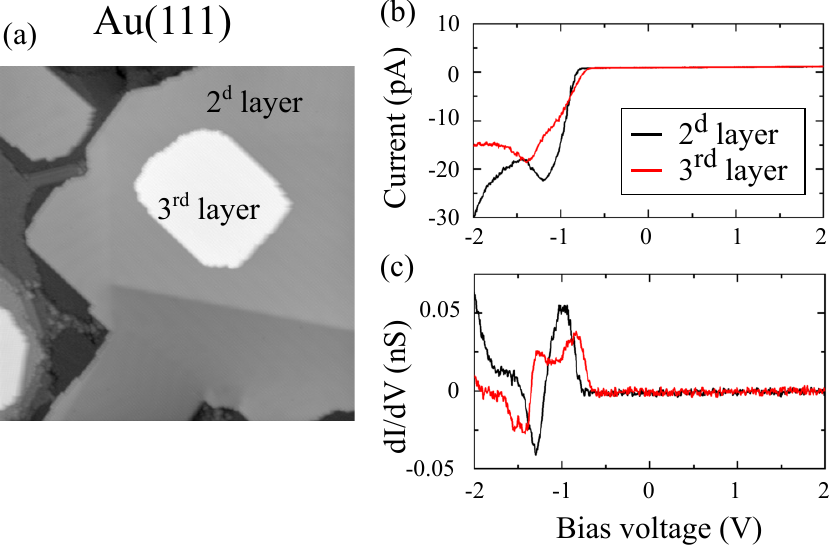}
		\caption{\label{STS_multi}
			(a) $150 \times 150$~nm$^2$ topographic STM image of FeMPz island up to 3~ML high absorbed on Au(111) ($-1.5$~V, 20~pA). (b) and (c) $I-V$ and $dI/dV$ curves recorded on the second (black curves) and the third layer (red curves). The presented curves are averaged over several molecules and measurements (2$^{\text{d}}$ layer: 20 curves on 4 molecules, 3$^{\text{rd}}$ layer: 15 curves on 3 molecules)}
	\end{figure}

	The electronic properties of the FeMPz molecules have also been investigated on several layers of molecules grown on Au(111). As illustrated by the STM topographic image of Fig.~\ref{STS_multi}a, at coverages above one monolayer, the molecular films follow a layer-by-layer growth with nicely defined edges. Fig.~\ref{STS_multi}b and c present the $I-V$ and $dI/dV$ curves measured on the second and third layer, respectively. One can clearly see that no state is available at positive bias voltage up to 2~V for the molecules in the second or third layer. Thus, the second and third layers cannot be imaged at positive bias voltages. At negative voltages, the NDR is still there but shifted to lower values as the molecular island height is increased. For molecules in the first monolayer, the $I-V$ curves acquired are ``noisy" due to the molecular switch under the STM tip (see Fig.~\ref{STS_mono}a). On the contrary, for the second and third layers, the $I-V$ curves are smooth, and the position of the NDR is reproducible from one molecule to the other. The maximum value of the NDR is at an energy of $-1.26 \pm 0.07$~eV for molecules in the second layer, and of $-1.42 \pm 0.10$~eV for molecules in the third layer.
	
We now focus on the different possible mechanisms at the origin of the observed NDR. The first thought is that it can arise from the molecular switching between the two spin states under the electric field applied between the tip and the metallic surface. In this case, the current measured as a function of the bias voltage switches between two extreme curves \cite{Miyamachi2012,gerhard2017electrically}. In the case of HOPG substrate, by measuring the $I -V$ curve in back and forth mode (i.e., the bias voltage is swept from $-2$~V to 2~V and then from 2~V to $-2$~V), abrupt switching is observed in the return curve (see Fig. S2 in SI). This brutal change is reproducible over various molecules and most probably originates from the spin state switching.

	Prior to and after the STS measurements on the molecules, the electronic state of the tip has been checked by acquiring spectra on bare Au(111) or bare Cu(111) to confirm the presence of the Shockley states (i.e., surface states). In some cases, after measurements over molecules, one can be picked up by the tip, giving rise to a specific signal. Fig. S3 in the SI reports two examples of such situations either on Cu(111) or on Au(111). In both cases, the NDR is observed but once for negative bias and once for positive bias. In the second case, one could think that the molecule is more strongly coupled to the tip than the surface, thus explaining the reversing of the $I-V$ curve \cite{Schull2009}. These experimental results tend to prove that the NDR signature observed is related to the intrinsic property of the FeMPz molecule.
	
\begin{figure}[!h]
		\centering
		\includegraphics[width=0.85\linewidth]{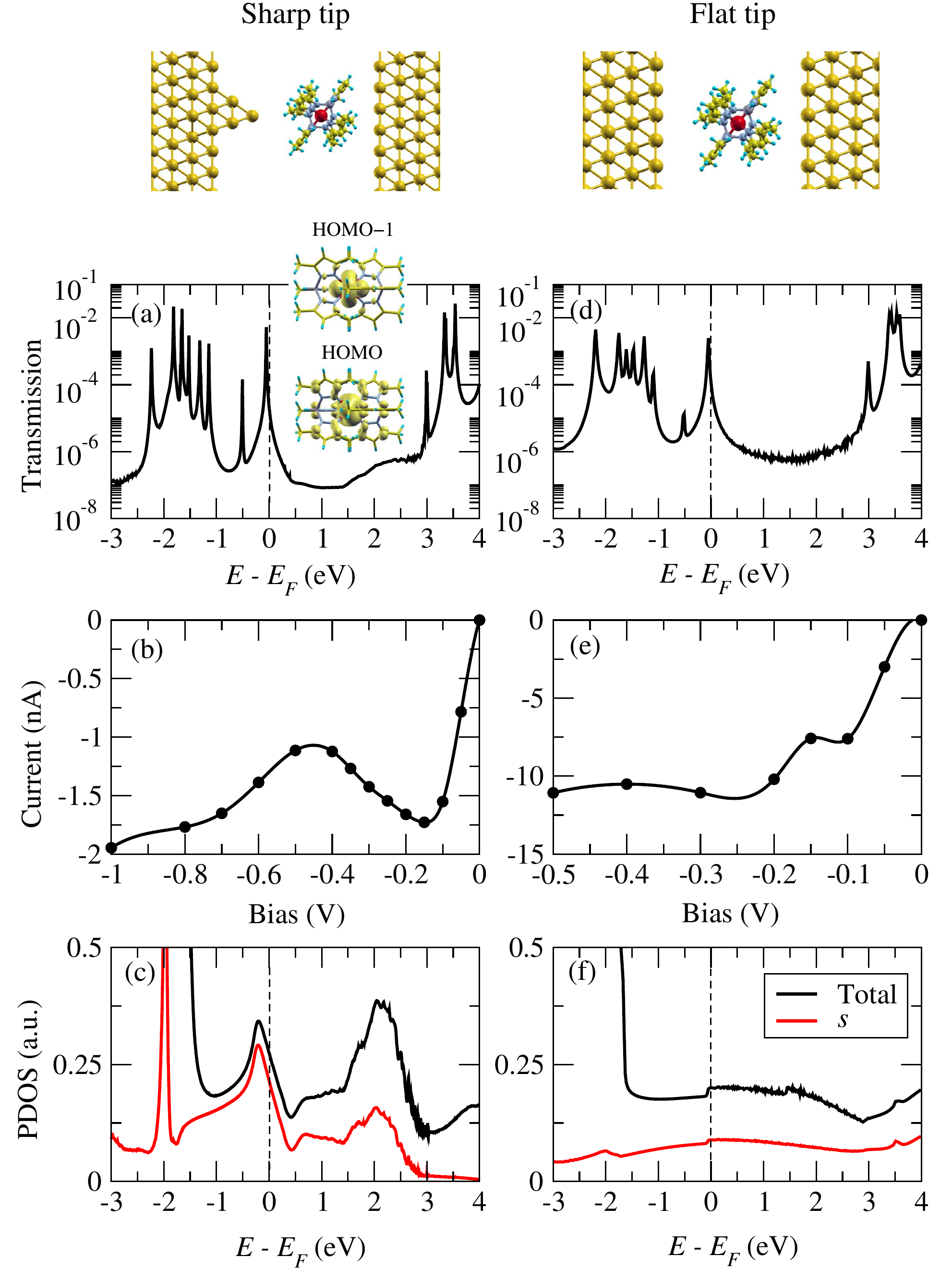}
		\caption{\label{i-v} 
NEGF+DFT results: (a) Zero-bias transmission as a function of energy for the low spin ($S$ = 0) configuration FeMPz molecular junction with an atomically sharp STM tip. (b) Current-voltage characteristics and (c) PDOS of the Au apex atom, where the black line denotes the total PDOS while the red line means the PDOS of the $s$ orbital. (d-f) The same as (a-c) but with a flat tip.}
	\end{figure}

	In order to explain such phenomena, we performed \textit{ab initio} quantum transport calculations for STM-based FeMPz molecular junctions as shown on the upper panels of Fig. \ref{i-v} for two considered geometries -- with sharp (on the left) or flat (on the right) STM tips. We used DFT at the PBE level including a van der Waals correction \cite{grimme2006semiempirical}, as implemented in the $\textsc{Siesta}$ \cite{soler2002siesta,garcia2020siesta} code. Subsequent electron transport calculations were carried out using \textsc{TranSiesta} \cite{Brandbyge-2002,papior2017improvements} and  post-processing codes \textsc{TBtrans} and \textsc{sisl} \cite{zerothi_sisl}, which employ the NEGF formalism combined with DFT (see SI for computational details).
	
    We start our discussion with the sharp tip geometry. From the zero-bias transmission function for the LS state ($S=0$), shown in Fig. \ref{i-v}a (in log scale), we observe a HOMO dominant transport behavior around the Fermi energy ($E_F$) with a large HOMO-LUMO gap of more than 3 eV. Due to the weak coupling between the HOMO-1 orbital and the Au surface, we observe a narrow transmission resonance at about 0.5 eV below $E_F$. This is also reflected in the spatial distributions shown on insets. The HOMO-1 is strongly localized on the Fe atom, originating from the Fe $d_{z^2}$ orbital (the $Z$ axis being directed along the molecule axis). The 2-fold degenerate HOMO, on the other hand, is more delocalized within the molecule (originating from Fe $d_{xz,yz}$ orbitals, which couple better to ligands) and therefore hybridizes much strongly with the surface, which results in a broadened transmission peak just below $E_F$. 
	
	Next, we consider the calculated current-voltage characteristics, $I-V$. The effect of applied bias, $V$, was included by shifting the levels of left (STM tip) and of right (substrate) electrodes symmetrically by $+V/2$ and $-V/2$, respectively. The Green’s functions of the scattering region (including the molecule and some Au atoms on both sides, shown on upper panels of Fig. \ref{i-v}) were then calculated self-consistently at each $V$. Interestingly, the simulated $I-V$ curve (Fig. \ref{i-v}b) reproduces the experimental observation reasonably well with a clear NDR signal observed for negative bias. The observed decrease in the current with voltage is a consequence of the lower amplitude of the HOMO-related transmission peak, as shown in Fig. S4 in SI. This peak, moreover, follows almost rigidly the Fermi level of the substrate when the bias is applied, reflecting a much stronger coupling of HOMO to the substrate compared to the tip. To get more insight, we plot in Fig. \ref{i-v}c the PDOS of the tip apex atom (black) and its $s$-orbital contribution (red). A strong energy-dependent behavior of the PDOS (mainly from $s$ orbital) near $E_F$ is clearly related to the atomically sharp geometry of the tip. In particular, a significant drop in the apex atom PDOS is observed right above the Fermi energy, which corresponds to the small negative bias where NDR appears. This indicates that NDR may indeed be related to the energy-dependent DOS of the tip apex atom rather than to the molecule itself. 
	
	To confirm this point, we further analyzed the coupling matrices, $\gamma$, of the HOMO orbital to both the surface and the tip (Fig. S5) which should be proportional to the corresponding PDOS. We find a dip-like feature in the $\gamma_{\text{Tip}}$ for negative $V$, correlating well with the drop of the tip atom PDOS discussed above, while the $\gamma_{\text{Surface}}$ is almost constant. The resonant transmission from the molecular level is given by $T=4\gamma_{\text{Tip}}\gamma_{\text{Surface}}/(\gamma_{\text{Tip}}+\gamma_{\text{Surface}})^2$. Therefore, one can conclude that the reduction of the HOMO-related transmission peak and, as a consequence, the NDR originate from the strong energy-dependent PDOS of the sharp STM tip. A similar effect was also reported in Ref. \cite{Xue1999}. 
	
	In order to eliminate the electronic effects of the tip, we have therefore performed transport calculations replacing the sharp STM tip with a flat surface. The transmission curve, Fig. \ref{i-v}d, looks very similar to the one for the sharp tip case, while the $I-V$, Fig. \ref{i-v}e, shows essentially different behavior. The current gradually increases with respect to the bias voltage; no NDR effect is observed. The PDOS analysis (Fig. \ref{i-v}f) also reveals rather constant PDOS around $E_F$ in this case.
	
	We note that the NDR effect is robust and well reproducible in our experiment, regardless of tip and substrate. This questions the tip-related mechanism presented above, rooted in the energy-dependent PDOS for a specific Au tip, pointing to a mechanism intrinsic to the molecule itself. Moreover, the mean-field-based DFT approach may not accurately capture strong correlation effects on SCO molecules coupled weakly to electrodes. We suggest, therefore, that the Coulomb blockade-like model introduced previously by Muralidharan and Datta \cite{Muralidharan2007} can also apply to the present case. The model is based on two kinds of molecular orbitals characterized by very different spatial localizations, exactly like in our case of HOMO (relatively delocalized) and HOMO-1 (strongly localized). The first one (with a large coupling to electrodes) provides the main transport channel (by holes) at a negative voltage. While the other one (weakly coupled) acts to block the transport when started, in its turn, to be emptied due to the Coulomb repulsion on the molecule strongly unfavoring a double hole configuration. 
	
\begin{figure}[!h]
		\centering
		\includegraphics[width=0.56\linewidth]{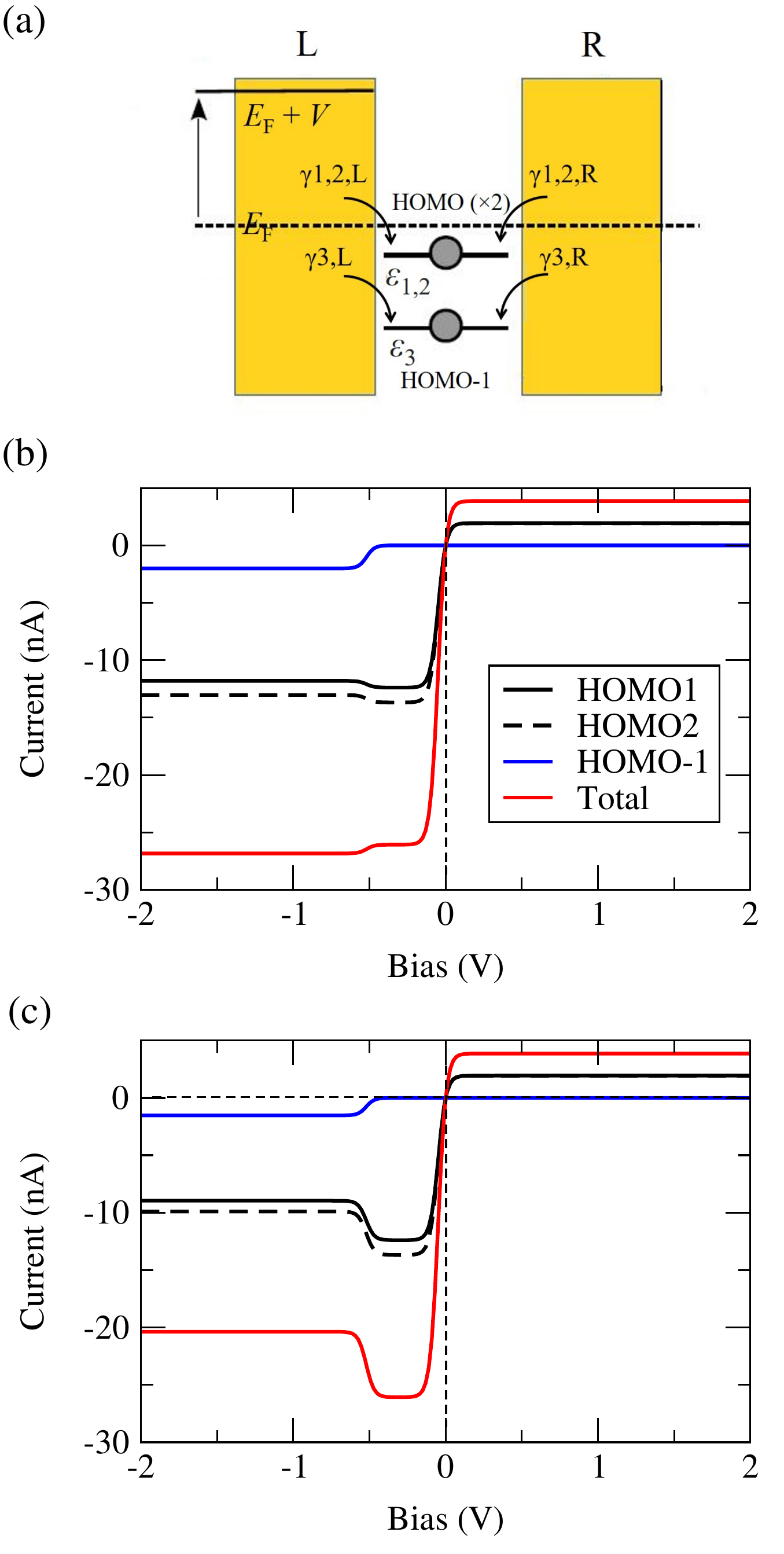}
		\caption{\label{model} Model results: (a) The effective three-level model based on the Coulomb blockade mechanism. Due to the weak coupling between the HOMO-1 and the electrodes, it plays the role of a "blocking orbital", which leads to NDR. The voltage is assumed to be applied fully to the left electrode (STM tip) simulating strong coupling of molecular levels to the surface (right electrode). (b) Simulated $I-V$ curve using parameters (in eV) extracted from DFT calculations: $\epsilon_{\text{1}}=-0.048$, $\epsilon_{\text{2}}=-0.051$, $\epsilon_{\text{3}}=-0.518$, $\gamma_{1,\text{L}}=6.995 \times 10^{-5}$, $\gamma_{1,\text{R}}=1.879 \times 10^{-3}$, $\gamma_{2,\text{L}}=7.725 \times 10^{-5}$, $\gamma_{2,\text{R}}=2.010 \times 10^{-3}$, $\gamma_{3,\text{L}}=1.190 \times 10^{-5}$, $\gamma_{3,\text{R}}=1.736 \times 10^{-4}$. Here, no NDR is observed. (c) The same as (b) but $\gamma_{3,\text{L}}=1.190 \times 10^{-5}$, $\gamma_{3,\text{R}}=2.380 \times 10^{-5}$. A clear NDR is found for negative bias.}
	\end{figure}

	The model (Fig. \ref{model}a) contains four (many-body) molecular states: the ground state of $N$ electrons, $\ket{0}$ (HOMO and HOMO-1 are occupied), and 3 states with $N-1$ electrons, $\ket{1}$, $\ket{2}$, and $\ket{3}$, corresponding to a hole on HOMO1, HOMO2, and HOMO-1 orbitals, respectively. They have energies $E_i(i=0,..3)$ and are characterized, in the sequential tunneling (weak coupling) transport regime, by their probabilities, satisfying the normalization condition, $\sum_{i=0,3}P_i=1$. In the out-of-equilibrium situation (applied voltage) the occupations obey the master equation: 
	\begin{equation}
	\begin{dcases}
	\frac{dP_0}{dt}&=-P_0\sum_{i=1,3}R_{0i}+
	\sum_{i=1,3}P_iR_{i0} \\
	\frac{dP_i}{dt}&=-P_iR_{i0}+P_0R_{0i},~~~~~i=1,2,3 
	\end{dcases}
	\label{eq-master}
	\end{equation}	
	where $R_{0i} = \sum_{\alpha = L,R} \Gamma_i^{\alpha}[1-f_{\alpha}(E_0-E_i)]$ describes the transition rate between the $N$ electron state $\ket{0}$ and $N-1$ electron states $\ket{i}$ with $\Gamma^{\alpha}$ being the level coupling strength to the electrode $\alpha$ and $f_{\alpha}(\epsilon)$ - the electrode's Fermi-Dirac distribution functions allowing the jump of an electron from the molecule to the electrode empty states (the zero energy is chosen to be the equilibrium Fermi energy, $E_F=0$). Opposite transition rates are given by $R_{i0} = \sum_{\alpha = L,R} \Gamma_i^{\alpha}f_{\alpha}(E_0-E_i)$ and are associated with the hopping of an electron from electrodes to the molecule. 
	
	In the steady state, the probabilities are constant in time and can be written explicitly:
	\begin{equation}
	\begin{dcases}
	P_0&=\frac{1}{1+\sum_{i=1,3}R_{0i}/R_{i0}}\\
	P_i&=P_0 \frac{R_{0i}}{R_{i0}},~~~~~i=1,2,3 
	\end{dcases}
	\label{eq-p}
	\end{equation}	
	
	The electron current from the left electrode to the molecule through the state $\ket{i}$ is then provided by:
	\begin{equation}
	I_{i}^L=-e^2/\hbar(R^L_{0i}P_0-R^L_{i0}P_i),
	\label{current}
	\end{equation}
	while the total current is obtained by summing all the contributions, $I^L=\sum_{i=1,3}I_{i}^L$.

	We can estimate the necessary parameters, energies $E_i$ and coupling constants $\Gamma_i^\alpha$, from DFT calculations, relating them to the energies and couplings of corresponding molecular orbitals, so that $E_0-E_{1,2}=\epsilon_{\rm1,2}, E_0-E_{3}=\epsilon_{\rm3}$ and $\Gamma_{1,2}^\alpha=\gamma_{\rm1,2}^\alpha, \Gamma_{3}^\alpha=\gamma_{\rm3}^\alpha$. Here, indexes 1, 2, 3 correspond to HOMO1, HOMO2, and HOMO-1, respectively. However, based on the set of parameters directly extracted from DFT, we could not reproduce the NDR (see Fig. \ref{model}b). 
	
	It is well known that the mean-filed DFT cannot accurately describe strong correlation effects due to the self-interaction error. Moreover, standard GGA functionals turn out to be not accurate enough to predict total energies of SCO molecule \cite{cirera2018benchmarking,cirera2020assessment}. Therefore, the parameters extracted from DFT are expected to be unreliable for strongly localized HOMO-1. Indeed, by setting $\gamma_{3,\text{L}} =0.5\gamma_{3,\text{R}}$, we could reproduce the experimental NDR feature reasonably well, as shown in Fig. \ref{model}c. In particular, when the negative voltage reaches the HOMO-1 level (at about $-0.6$ V), the three levels starts to be emptied simultaneously. However, due to the weak coupling of the HOMO-1 and electrodes, the hole stays on HOMO-1 much longer compared to HOMO orbitals. On the other side, the double-hole occupation (in all the HOMOs and HOMO-1) is strongly suppressed due to a substantial charging energy on the molecule (which is encoded in the normalization condition of occupations of four molecular states involved in the model). 
	
	As a result, the occupancies of HOMO-related states are reduced, leading to a decrease in the associated currents (black solid and dotted lines). Therefore, the strongly localized HOMO-1 orbital acts as a "blocking state": its small contribution to the total current is not enough, and the total current start to decrease, resulting in NDR. Finally, it should be noted that the DFT parameters (which did not work straightforwardly to produce NDR) were extracted from the low spin (LS, $S=0$) state, while the NDR is more often observed experimentally for the high spin (HS, $S=2$) state. The DFT calculations for the magnetic HS state turned out to be quite delicate for the \textsc{Transiesta} code (to find correct level positions for frontier molecular orbitals), especially when the Coulomb+$U$ correction (which is an unknown parameter in general) is applied. Therefore, we did not manage to obtain reliable parameters for this case. On the other hand, the proposed arguments for NDR should remain valid also for the HS state due to the fact that the same orbitals are involved in the transport:  HOMO is mainly from $d_{xz,yz}$ (delocalized) while the HOMO-1 is from $d_{z^2}$ (strongly localized). This is confirmed by DFT+$U$ calculations for free-standing SCO of the HS (see Fig. S6 in SI). As a result, we can speculate that the model parameters, if extracted from the HS case, could fit better with those reproducing the NDR.
	
	To conclude, we report on a pronounced NDR effect from Fe$^{\text{II}}$ SCO molecules deposited on surfaces, highly robust to substrate materials (Au, Cu, or HOPG), temperature, and the number of SCO layers. A generic theoretical model based on the Coulomb blockade mechanism captures the experimental findings pretty well, tracing the origin of NDR to strongly localized HOMO-1 and rather delocalized HOMO orbitals. We also discuss that the mean-field-based DFT+NEGF approach fails to capture the intrinsic NDR effect due to presumably strong correlation effects on the SCO molecule. We believe that our findings and proposed physical mechanism bear a rather general character and can apply to other molecular systems with similar electronic properties.
	
	\noindent{\bf Acknowledgment}
	This project received funding from the European Union’s Horizon 2020 Research and Innovation Programme under grant agreement no. 766726. The Center for Nanostructured Graphene (CNG) is sponsored by the Danish National Research Foundation (DNRF103). Part of the calculations was done using HPC resources from CALMIP (Grant 2022-[P21008]). The authors would like to acknowledge Marie-Laure Boillot and Talal Mallah for providing the spin-crossover molecules and Cyril Chacon for experimental support. 

\begin{suppinfo}
	The Supporting Information is available free of charge at \url{https://pubs.acs.org/}. Experimental and computational details, spectroscopy of FeMPz molecules on Au(111) at 78 K and on HOPG for switching, spectroscopy of FeMPz molecules adsorbed on the tip, bias-dependent transmission functions, tip and surface coupling strength projected on HOMO, the DFT+$U$ results for free standing SCO.
\end{suppinfo}

\bibliography{Bibliomanuscript}

\newpage	
	
\end{document}